# PREDICTING MASSIVE HELIUM-3 RELEASE FROM METAL TRITIDES USING SIMPLE MECHANICAL MODELING.

Bérengère Evin[1*], Dorian Gaboriau[1], Mathieu Segard[1], Sylvain Challet[1], Arnaud Fabre[1], Stéphanie Thiébaut[1]
[1] CEA, DAM, Valduc, F-21120 Is-Sur-Tille, France
ORCID : 0000-0002-8305-1203

**ABSTRACT**. This letter is presenting a simple but effective mechanism that explains why ,during tritium aging, metal tritides retain most helium-3 for years and then suddenly release massive amount. The mechanism is based on the hypothesis that dislocations blocking could explain the sudden change of behavior. The modeling of this phenomenon combine a mechanical and microstructural approach. The calculations made with this mechanism fit all the aging data acquired on aged palladium tritide.

## I. INTRODUCTION.

Metal tritides are used to reversibly store tritium, the radioactive isotope of hydrogen, at low pressure as they ensure safety and compactness. As tritium decays ($t_{1/2}$=12.3 years), helium-3 is produced and is mainly retained in metallic tritides as nanobubbles (only a slow weeping of a few percent is observed). After a few years, most helium-3 is massively released in the gas phase. Helium-3 desorption threshold is of great interest. Indeed, beyond it pressure rises quickly in the container. It is therefore of utmost importance in order to design safe tritium storage systems. This phenomenon is observed in many metal tritides at ~2 years for $UT_x$ [1], ~9 years for $PdT_x$ [2], ~6 years for $TiT_x$ [3], ~5 years for $ZrT_x$ [4].

Helium-3 behavior in metal tritides is broadly documented as a two-step process, with first helium-3 predominant retention, and then helium-3 massive desorption. Despite the fact that helium-3 in metals has been studied for a long time [5], the mechanism underlying the threshold between retention and desorption remains unclear. Moreover, available experimental data is scarce, since it takes years/dozens of years to age a material under tritium long enough to experimentally study this phenomenon.

A semi-empirical modeling was previously proposed for $PdT_x$, which succeeded in modeling helium-3 behavior in both retention and desorption phase [6]. However it does not explain why metal tritides "suddenly" shift from retention to desorption.

In this letter, a simple yet effective mechanism is proposed for $PdT_x$. It predicts helium-3 behavior in palladium tritide and explains why $PdT_x$ is shifting from retention to desorption. This mechanism is based on the idea that helium bubbles could be considered as solid-state precipitates which increase tritide hardness, until it reaches a maximum. It enables the establishment of a criterion, based on dislocations mobility, that defines the threshold between helium retention and helium desorption.

The dislocation model considers that, as long as dislocations can move, helium can be retained in the tritide as nanobubbles can grow. When dislocations can no longer move, nanobubbles cannot grow. Maximum hardness is reached, plastic deformation is no longer possible and newly created helium is released.

The mechanism was implanted in a previously published continuum mechanics modeling that predicts nanobubbles growth in palladium [7]. The modeling successfully matches all experimental measurements made on palladium tritide.

## II. METHODS

### A. Analysis of an old Pd sample

The first analysis of a palladium powder aged under tritium long enough to reach the desorption phase was published in 2023 [6]. This sample was aged over 20 years, reaching a theoretical He/Pd ratio of 0.46 and a measured one of 0.20. The theoretical He/Pd ratio (He/Pd$_{th}$) is the ratio that would be reached if all helium was retained in the palladium. The real He/Pd ratio (He/Pd$_r$) takes into account helium release in the gas phase during aging. The sample provides valuable data on desorption mechanism and helium-3 behavior.

Its analysis shows that bubble (or cavity) density, bubble (or cavity) size and helium bubble volume in palladium are similar to those measured on younger samples. To summarize, microstructure is not fundamentally changed after massive $^3$He desorption as illustrated by images from Transmission Electron Microscopy (TEM) in Figure 1.

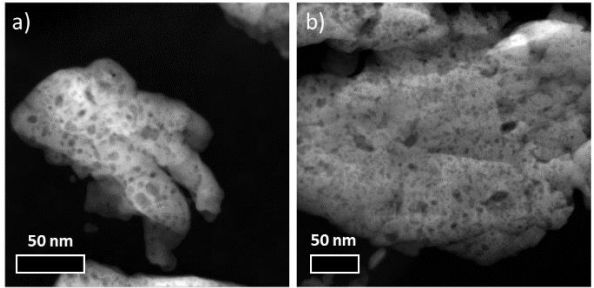

*Figure 1 – High Angular Annular Dark Field images of palladium powder aged 6.5 years (He/Pd$_r$=0.23) (left) and over 20 years (He/Pd$_{th}$=0.46) (right) under tritium.*

Since microstructure remains similar after helium desorption, it was assumed that the bubbles could be empty. However, Nuclear Magnetic Resonance (NMR) and Electron Energy-Loss Spectroscopy (EELS) measurements agree on bubble pressures that are of a few GPa, like few-years-old samples.

To conclude, this sample was unexpectedly similar compared to its younger counterparts. This leads to the idea that desorption mechanism does not involve measurable structural change nor massive emptying of the bubbles.

These experimental observations have led to the semi-empirical modeling published in 2023 [6] where the desorption threshold is imposed. It succeeds in modeling $^3$He behavior but does not explain the underlying mechanism, which is the point of the present work.

**B. Mechanism idea**

The first observation is that helium bubbles in palladium tritide are so pressured that they could no longer behave as gas bubbles in metals [6], [8]. These nanobubbles behave more like solid-state precipitates. It could be compared to precipitation hardening which is used to increase metal hardness using an impurity phase to impede the movements of dislocations. In steel, helium-induced hardening was measured in tritium-aged steel-components [9].

Second observation is that most metal tritides behave similarly regarding helium-3 retention behavior, with $^3$He retention then massive desorption [1], [2], [4]. This suggests that many metals, regardless their differences, face some breaking point in helium-3 retention.

Last important point is that the beginning of helium-3 desorption is correlated with the end of palladium tritide swelling. This suggests that the material can no longer withstand plastic deformation. This means that dislocations (or defects) can no longer move.

Immobilization of dislocations can also be understood as a rise in the tritide hardness. In fact, the threshold between retention and desorption can be interpreted as reaching the maximum hardness of the metal tritide.

At this point, the difficulty is that there is no measurement of metal tritide hardness during aging. On a wider scale, there are few available data on mechanical properties during tritium aging. However, when looking at the way to increase hardness using precipitation hardening in metallurgy, there are indications that apply to helium-3 nanobubbles [9]. For example, it is possible to calculate the size and spacing of the precipitates that best increase the hardness of a metal.

These calculations are actually based on dislocations motion mechanisms, Orowan's bypass [10] and Friedel's cutting [11]. Orowan and Friedel strains (Eq. (1) and (2)) are the mechanical strain to overcome to move dislocations by cutting or surrounding precipitates:

$$\sigma_O = \frac{0.7\mu b \sqrt{f_V}}{r} \quad (1)$$

$$\sigma_F = \sqrt{\frac{3\alpha^3}{2\pi}} \mu \sqrt{\frac{f_V r}{b}} \quad (2)$$

$$\Delta = \sigma_O - \sigma_F \quad (3)$$

where μ is the shear modulus (in GPa), b is the Burger's vector (in nm), $f_V$ is the volume fraction of precipitates (no dimension), r is the precipitate radius (in nm) and α is the interaction constant (no dimension) [12], [13].

Equation (1) shows that Orowan strain rises when the precipitate radius decreases. Orowan mechanism is favored for materials containing large and wide-spaced precipitates. On the contrary, equation (2) states that Friedel strain rises as precipitate radius increases. Friedel mechanism is favored in materials containing small and close precipitates.

If precipitates are too close for Orowan mechanism and too large for Friedel mechanisms both strains are equal [14]. In this case, none mechanism is favored and dislocations movements are blocked. This point is reached when Δ (Eq. (3)) is minimized.

Equalizing $\sigma_O$ and $\sigma_F$ gives the precipitate radius for which none mechanism is favored, thus dislocations are blocked. A small order-of-magnitude calculation leads to radius r≈1.4 nm for PdT$_x$. This value is corroborated by TEM measurements which give bubbles diameters distributions centered between 2 and 3 nm for aged Pd samples [15].

**C. Implementing mechanism in modeling**

*Contact author: berengere.evin@cea.fr

The bubble growth mechanism is already published in Ref [7]. It is based on Continuum Mechanics. The mechanisms that predict dislocations movements were added to it.

The modeling considers a helium-3 bubble surrounded by a metal tritide matrix. As the bubble grows, the REV (Representative Elementary Volume) grows following Eq. (4) and (5). The metal matrix plasticizes and helium content in the bubble rises. The initial size of the REV is based on bubble density (see Eq.(6)):

$$\sigma = \sigma_1 + k\dot{\varepsilon}^m \quad (4)$$
$$\sigma_1 = \sigma_\infty - (\sigma_\infty - \sigma_0)e^{-\xi\varepsilon} \quad (5)$$
$$REV \to \frac{4}{3}\pi R_0^3 dB = c \quad (6)$$
$$NHe_{real} = (1-S)\frac{stoe.\rho_{Pd}.N_a}{M_{Pd}}(1 - e^{-\lambda t}) \quad (7)$$
$$NHe_{theo} = \frac{stoe.\rho_{Pd}.N_a}{M_{Pd}}(1 - e^{-\lambda t}) \quad (8)$$
$$NHe_{des} = NHe_{theo} - NHe_{real} \quad (9)$$

where $\sigma$ is stress, $\sigma_0$ yield strength, k consistency, $\xi$ hardening, m deformation sensitivity, $\varepsilon$ strain, $\sigma_\infty$ yield stress, $R_0$ the initial REV radius, dB the bubble density, c compactness. $NHe_{theo}$, $NHe_{real}$ and $NHe_{des}$ are respectively helium produced during aging, helium stored in tritide and helium released to the gas phase. S helium slow weeping, stoe tritide stoichiometry, $\rho_{Pd}$ Pd density, $N_a$ Avogadro's number, $M_{Pd}$ molar mass of Pd, $\lambda$ tritium constant decay, t time.

At each calculation step (≈1 day), the bubble growth modeling takes into account newly created helium-3 which is integrated in the bubble. Orowan and Friedel strains calculations were added to it. As long as Orowan and Friedel mechanisms are competing, helium-3 is stored in the palladium tritide (in the nanobubble). If one of the mechanisms is favored it means that dislocations can move, which implies that palladium tritide can sustain plastic deformation and that maximum hardness is not reached.

When Orowan and Friedel strains are equal (Δ defined in Equation (3) close to zero), dislocations can no longer move. Maximum hardness is reached. Its means that palladium tritide can no longer withstand plastic deformation (which is necessary for bubble nucleation and growth). Consequently, since no new bubble nor existing one can accommodate more helium-3, newly created helium-3 is released to the gas phase.

This point marks the threshold between helium retention and helium desorption. Helium-3 previously trapped in bubbles in palladium tritide remains in it. No other condition is added to threshold this bubble growth modeling.

## III. RESULTS

In this section the modeling results are presented. The initialization inputs of the models are $r_0$=0.1 nm, $R_0$=2.5 nm, slow helium-3 weeping during aging S=4%, strain hardening β=360, yield strength $\sigma_1$=170 MPa, tensile strength $\sigma_\infty$=1200 MPa, Burger's vector b=0.27 nm [16], α=0.2, shear modulus μ=47 GPa [17].

Modeling outputs are the bubble radius, the bubble pressure, the $He/Pd_r$ and $He/Pd_{th}$ ratio. Each output is compared with experimental measurements made on palladium powders aged under tritium which are available and detailed in Ref [6].

The new dislocations modeling (DM) proposed in this letter is compared to the previous semi-empirical modeling (SEM) from [6]. The main difference between these two modelings is the condition to enter desorption phase. In the SEM, desorption phase is imposed when helium density in the bubble reaches 150 $He/nm^3$ and bubble volume 30%. In DM the material enters desorption phase when dislocations are blocked ($\sigma_O = \sigma_F$).

Figure 2 is the evolution of bubble radius over time.

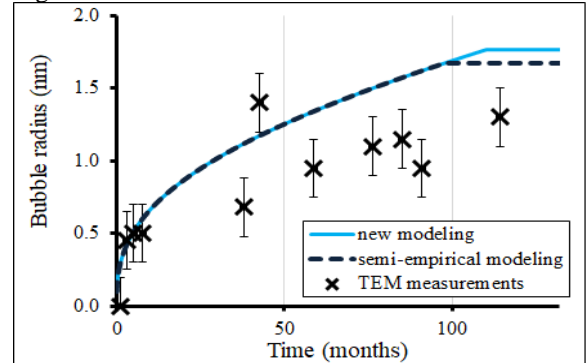

*Figure 2 – Bubble radius evolution over time. New dislocations modeling (solid line) is compared to previous semi-empirical modeling (dotted line) and experimental measurements of TEM and STEM tomography [18] (cross).*

Bubble size increases with time as shown in Figure 2. As bubble growth mechanisms are the same in both modeling, the evolution of bubble size with aging time is similar. As massive desorption is reached, the bubble stops growing around 1.7 nm. The new dislocations modeling successfully retrieves helium desorption threshold that was imposed in the semi-empirical one.

In the previous semi-empirical modeling, the desorption threshold occurs at 100 month of aging, the new one, proposed in this letter, occurs at 110 month, in agreement with what is observed experimentally (108 months [2]).

*Contact author: berengere.evin@cea.fr

Figure 3 shows the bubble pressure over time.

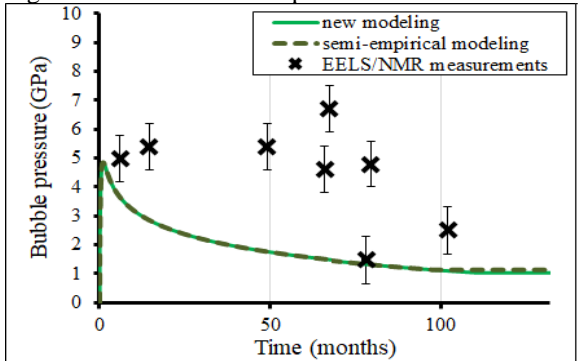

*Figure 3 – Bubble pressure as a function of time. New dislocations modeling (solid line) is compared to previous semi-empirical modeling (dotted line) and experimental measurements of EELS [8] and NMR (cross).*

Bubble pressure is of a few GPa, but the modelings under-estimate the pressure. This is probably an effect of the helium Equation Of State implanted in the model. The helium desorption occurs for a bubble pressure between 1 and 2 GPa.

Figure 4 presents the evolution of He/$Pd_r$ ratio over the He/$Pd_{th}$ ratio.

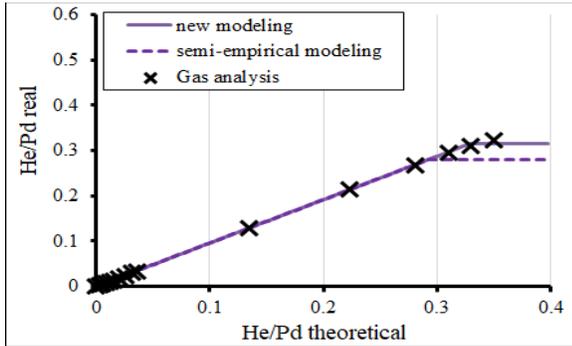

*Figure 4 – He/Pd ratio evolution as a function of the theoretical He/Pd (no helium release). New dislocations modeling (solid line) is compared to semi-empirical modeling (dotted line) and experimental measurements of gas analysis (cross) [18].*

The new modeling predicts that the threshold of helium retention leading to desorption occurs for a He/Pd ratio of 0.33 which is in agreement with what can be measured experimentally.

## IV. DISCUSSION

The outputs of the new dislocations modeling suitably fit the experimental data acquired on aged palladium powders. The simple mechanism based on dislocations movements implemented in the bubble growth modeling successfully retrieves the helium desorption threshold.

This simple mechanistic modeling for $PdT_x$ is interesting as it does not need much data on aged materials to run. This new addition does not impact the bubble growth step (< 9 years of aging).

The previous modeling succeeded in reproducing helium-3 behavior in tritides but needed helium retention data (threshold bubble density and volume) to retrieve helium desorption threshold. The new one is not semi-empirical. This new modeling is a great improvement as it is able to predict desorption behavior based on dislocation movements.

To our knowledge, no mechanism has been proposed to explain helium massive desorption. The modeling proposed in this letter based on dislocation movements' mechanisms succeeds in predicting helium desorption. The calculation is simple but seems effective.

## V. CONCLUSIONS

Considering that helium-3 "bubbles" in tritides behave as solid-state precipitates successfully leads to a retention-desorption modeling that fits all experimental observations on palladium.

The continuum mechanics modeling is based on the fact that retention is possible as long as dislocations are mobile. When dislocation mobility is mechanically hindered ($\sigma_F = \sigma_O$), the tritide reaches its maximum hardness and newly created helium-3 is no longer retained. This suggests that helium-3 retention in tritides can be tuned by bubble fraction, bubble size, but also by the metallurgic state of the host metal.

The simple modeling of the phenomenon needs some experimental data to be run, but even with some order of magnitude, it gives a fairly accurate idea of the tritide's behavior over time regarding helium-3 desorption. Moreover, for materials that faces tritium, helium-3 retention can be predicted with the initial material properties without having to conduct long-term aging experimental studies. Helium-3 retention and desorption has no longer to be suffered but can be anticipated.

Future works will consists in applying this modeling to other tritide.

## ACKNOWLEDGMENTS
This work was supported by France's Atomic Energy Commission (C.E.A.). The authors acknowledge CEA for financial support.

*Contact author: berengere.evin@cea.fr

*Contact author: berengere.evin@cea.fr